# Hydrogen Diffusion in Silicon – An ab initio Study of Hydrogen Kinetic Properties in Silicon

Liviu Bilteanu[1,2]*, Mathias Posselt[3], Jean-Paul Crocombette[1]

1 Services des Recherches de Métallurgie Physique, CEA Saclay, 91191 Gif-sur-Yvette, France.

2 Laboratoire de Physique des solides UMR 8502, Université Paris Sud 11, 91405 Orsay, France,

3 Forschungszentrum Dresden-Rossendorf (FZD), PF 510119, D-01314 Dresden, Germany.

Contact Author: Liviu Bilteanu, liviu.bilteanu@u-psud.fr, Tel. +33 1 69 15 69 37

Abstract

In this paper we present kinetic properties such as migration and decomposition barriers of hydrogen defects in silicon calculated by Density Functional Theory (DFT) based methods. We study the following defects: H atoms (H) and ions ($H^+$, $H^-$), hydrogen molecules ($H_2$) and hydrogen complexes ($H_2^*$). Our results show that the dominating migration species are H ions, their charge-independent migration barrier being 0.46 eV in excellent agreement with experimental value[1]. At higher temperatures $H_2$ and $H_2^*$ decompose with barriers of 1.2 and 1.5 eV respectively, while the migration barrier of $H_2$ is 2.2 eV. Hence, we show that, contrary to what has been experimentally implied[2,3] previously, in silicon, hydrogen molecules do not participate to the hydrogen transport through diffusion.

# 1   Introduction

Hydrogen ion implantation in semiconductors has revealed the creation of large bidimensional defects called platelets, a phenomenon relevant for technological purposes[4-7]. Several studies have been dedicated to describe the growth[8-14] or the structure[15-18] of these defects. Understanding the platelets formation is conditioned by a clear description of the early stages of the hydrogen into these extended defects. This means that one needs an atomic scale description of the hydrogen diffusion through mechanisms of hydrogen migration between two stability sites.

Density Functional Theory (DFT) has proved to be a fruitful computational tool for the investigation of the hydrogen related defects in silicon. Through the pioneering work of Van de Walle and coworkers[19-23] one has been able to describe quantitatively the energetics of hydrogen atoms and molecules in silicon. The main results that we retain from these works:

- the hydrogen may bear charge especially in atomic states
- the hydrogen is more stable under diatomic forms rather than in atomic states

This paper is dedicated to the study of charge effects on hydrogen stability in silicon and especially on the kinetic properties of two types of hydrogen configurations in silicon: monatomic and diatomic configurations. By kinetic properties we understand the migration barriers of monatomic and diatomic hydrogen and the decomposition barriers of diatomic hydrogen.

The rest of the paper is divided in four sections. The next section, Section 2 is dedicated to the description of the computational details of our calculations. Section 3 is dedicated to the presentation of results. In this section we shall show that the dominating migrating hydrogen species are the charged hydrogen atoms and we shall present their migration barriers that are in excellent quantitative agreement with the experimental literature. Section 4 is dedicated to the discussion our results in the light of the experimental findings and in the light of earlier computational results. In the final section we draw up the conclusions of our investigation.

# 2  Method

All the DFT calculations presented in this section have been performed with the SIESTA code[24] using norm-conserving pseudopotentials and bases of numerical atomic orbitals. The exchange-correlation has been treated within the Generalized Gradient Approximation (GGA)[25]. The pseudopotentials were provided with the SIESTA package and the employed basis sets were of non-polarised double-zeta type. Their radial cut-off values are 7.0 Bohr for Si and 8.0 Bohr for H[24, 26]. The Si basis has been constructed in order to provide the correct electronic properties of the perfect bulk silicon. The main convergence tests that have been performed are the dependence of the formation energy of silicon selfdefects on the supercell size and the k-space sampling. Following these convergence tests, the calculations have been carried out using 216 atoms 3×3×3 supercells. The lattice parameter is 5.431 Å. This value is identical to the experimental value. The k-space is sampled by 8 points (that is 2×2×2 in a Monkhorst-Pack scheme). The cut-off energy used in the calculations is 150 Ry. The good agreement between our results and the recent ones on the same type of defects in comparable conditions[27, 28] ensure the correctness of this computational configuration. Indeed, we have obtained a formation energy of a neutral vacancy of 3.57 eV as compared to 3.36 eV[27] and 3.69 eV[28] in literature, while the formation energy for a neutral tetrahedral interstitial was 3.82 eV (compared to 4.06 eV[28]). All our calculations are performed allowing the supercell volume and shape relaxation, the maximum force being required to be less than 0.04 eV/Å.

The first step in our investigation is to check the stability of the hydrogen related structures. Actually this step was absolutely necessary in order to get the initial and the final states for the later calculations of migration and decomposition barriers. Within the first step the supercell is relaxed under the conditions described above and one obtains two types of results

- total energies of the configurations that we took into account;
- the relaxed positions of all atoms in the supercell

The total energies are used to calculate the formation energies and relative stabilities of defects. The calculation formulas/schemes for these quantities will be provided for each type of defects in the Results section. The comparison with the literature of these quantities (we make in the Discussion section) will provide a validation of the chosen numerical scheme.

To calculate the migration barriers we have used the Drag method[29]. Prior to making a defect-containing system evolve from the state X to state Y (where X and Y are two vectors containing all the positions of the atoms in the simulation box initial and final points respectively). These states have been obtained before by simple relaxation of the systems containing the defect in the initial (X) and the final (Y) positions.

We construct intermediary independent images of the system by linear combination between the X and Y vectors. Each image can be represented formally by the vector $R_\lambda = X + (1-\lambda)Y$ ($\lambda \in [0, 1]$). Then we relax the system in the hyperplan perpendicular on the hyperline XY in the point $R_\lambda$. The values of the system energy after full relaxation in the hyperplan are used to construct the migration barrier. In contrast with the Nudged Elastic Band, within the Drag method the intermediary images of the system are independent, this meaning that the relaxation of a system in $R_\lambda$ is independent of the relaxation of the system in the $R_{\lambda-1}$ and $R_{\lambda+1}$. This leads to a significant gain in the computation time, without sacrificing to much the accuracy of the barrier calculations.

Forcing the system to evolve between two equivalent states provides the energy barrier of such process that can be either a migration or a decomposition process.

## 3 Results

### 3.1 Atoms

#### 3.1.1 Stability of atomic structures

Hydrogen atoms are the simplest defects in silicon and in this subsection we shall present the results on their stability and their migration between two equivalent sites.

We have tested the stability of hydrogen atoms in three charge states (±1, 0) in various sites (see Figure 1). In Figure 1 we have aligned on the [111] directions some relevant stability and meta-stability sites of H in Si as follows: the bond-centre (BC) site – the midpoint of a Si-Si bond; the anti-bonding (AB) site – the mirror point of the BC-site with respect to one Si atom of the bond; the tetrahedral (T) site - the mirror point of one Si atom with respect to the nearest neighbour Si atom; the C-site – the midpoint between the next-nearest neighbour Si atoms; the M-site – the midpoint between nearest neighbour C-sites and finally; the high-symmetry hexagonal (Hex) site – the midpoint to the next nearest neighbour C-sites or the midpoint between two T sites.

The physical quantity used to describe the stability of various atomic defects is the formation energy (or solution energy) of hydrogen in silicon. This quantity is written as:

(1) $$E_F = E_{H\ in\ Si} - E_{Si\ bulk} - E_{H\ free} + qE_{Fermi}$$

with q = ±1. The solution energy contains four terms. The first three terms obtained directly from the SIESTA code output are: $E_{H\ in\ Si}$, the energy of a system containing one hydrogen atom, $E_{Si\ bulk}$, bulk energy in a defect free box and $E_{H\ free}$, the energy of the free hydrogen atom.

The calculation of $E_F$ for each pair site-charge allowed the construction of stability diagrams in all three charge series: neutral, positive and negative. The most stable sites for hydrogen atoms are the following ones: BC for H$^0$, BC for H$^+$ and AB for H$^-$ (Figure 2).

### 3.1.2 Negative U

When plotting the solution energy (1) against the Fermi level (Figure 3) one observes the preference of hydrogen atoms for charged states rather than for the neutral state for all values of the Fermi level ($E_{Fermi}$) in the band gap. Such a preference, exhibited also by hydrogen in other semiconducting materials (ZnS, ZnSe, GaN etc.) is called the negative U effect.

The charge sign depends on the position of the Fermi level within the gap. More exactly within the gap going from the valence band maximum (VBM) to the conduction band minimum (CBM) instead of the standard order (+/0), (+/-) and (0/-) of the charge transition levels, in the case of the negative-U effect the order is (0/-), (+/-) and (+/0). The value of the U constant determined from Figure 3 is 0.32 eV in perfect agreement with the experimental value of 0.36 eV[30, 31]. The value of the *U* constant is the double of the distance between the cross point between the formation energy of H$^+$ and H$^-$ and the horizontal line representing the formation energy of the neutral hydrogen, $H^0$. This value can be interpreted as the energy gain for the following process:

(2)  $$2H^0 \rightarrow H^+ + H^-$$

For a specific position within the gap (denoted $E_{Fermi}^{\pm}$) the hydrogen atoms exist in both charged states in equal quantities. According to our calculations, this position is very close to the midgap:

(3)  $$E_{Fermi}^{\pm} = E_{VBM} + 0.54 \text{ eV}$$

The position of the Fermi level within the gap depends on the hydrogen concentration in the system. We distinguish two cases:

1. The case of small H dilutions – in which the position of the Fermi level is independent on H concentration, but depends on the temperature and on the concentration of other eventual doping atoms in higher dilutions.

2. The case of high H concentrations in which hydrogen atoms become themselves doping atoms. In this case there is an auto-coherent dependence between the Fermi level and the charged hydrogen defects.

In the case of high-dose implantation the number of vacancies produced is considerable. The second case we mention before is valid if the effect of vacancies is annihilated by hydrogen saturation of the vacancy dangling bonds.

### 3.1.3 Migration of H atoms

In order to produce platelets one has to implant hydrogen in atomic percentages that can be very high. We fall hence into the second case: the hydrogen is the dominating species and controls the position of the Fermi level. The position of the Fermi level in the gap corresponds to the charge equilibrium of hydrogen atoms (the crossing point of formation energy lines of $H^+$ and $H^-$, in Figure 3).

$H^0$ is less stable than $H^+$ and $H^-$ so for the calculation of kinetic barriers we consider the migration of $H^+$ between two BC sites, the migration of $H^-$ between AB sites and the migration $H_2$ between two T sites.

We have therefore considered the starting and the ending points to be equivalent sites: BC sites for $H^+$ and AB sites for $H^-$. In the case of $H^+$ the considered sites are the first order neighbors BC sites (Figure 4a). In the case of the negative hydrogen ($H^-$) the migration from one AB site to the first order nearest-neighbour (NN) AB site will actually correspond to a rotation around the same Si atom. In this case, however the migration distance is rather small and the barrier is very low. A migration process corresponding to a larger distance would be the displacement of $H^-$ from an AB site to its second order NN AB site (Figure 4b).

Figure 5 shows that for these two $H^\pm$ migration processes the barrier is the same, 0.46 eV in a striking agreement with to the experimental value of 0.48 eV obtained for atomic hydrogen by van Wieringen and Warmoltz[1]. This shows that the migration of atomic hydrogen is charge independent.

Furthermore, when calculating the migration of a positive $H^+$ from a BC site to its $2^{nd}$ nearest neighbour it has been observed that the barrier is in fact composed by two successive hopping of ~0.5 eV each from a BC site to the first order NN BC site (Figure 6). This means that the migration process we have described for $H^+$ is an elementary migration process in silicon.

## 3.2 Diatomic Structures

### 3.2.1 Stability of di-atomic structures

Like in vacuum (free hydrogen atoms gas), in silicon hydrogen atoms tend to agglomerate at room temperature in order to form diatomic structures. In order to compare the diatomic configurations to the monatomic configurations (Figure 2), we have used the formation energy per hydrogen atom that is the total formation energy divided by two (the number of atoms within the diatomic structures). The total formation energy can be expressed, with respect to an exterior hydrogen reference (such as a free hydrogen atom or molecule) or with respect to a reference representing the hydrogen inside the silicon lattice. In the second case the most used reference[19-23, 32, 33] is the neutral hydrogen atom in the most stable configuration (BC site). Since we compare various defects inside the Si bulk it seems natural to choose as hydrogen reference the energy of the neutral hydrogen atom in the BC site. In this case, the formula for the formation energy per hydrogen atom reads as follows:

(4) $$E_F^{H_2} = \frac{1}{2}\left(E_{H_2 \text{ in } Si} + E_{Si \text{ bulk}} - 2E_{H \text{ in } Si} + qE_{Fermi}\right)$$

Whereas $E_{H_2 \text{ in } Si}$ is the energy of a dimer in Si, while the last three terms have been explained in the previous section. This energy corresponds actually to two processes: the formation of a diatomic structure followed by the loss or the gain of $|q|$ electrons.

In this work we shall present results on neutral dimer only due to the fact that the energies we have obtained for the charged ($q = \pm 1, \pm 2$) dimer structures is up to 4.0 eV higher with respect to the neutral ones. So we have excluded the possibility of having charged dimmers inside the Si crystal.

Several dimmer configurations have been investigated within this study, however only three configurations proved to be stable: the interstitial $H_2$ molecule in T site (oriented within [111] direction) and in the Hex sites (perpendicular to the hexagonal plan) and a diatomic complex, $H_2^*$ formed by two hydrogen atoms one in AB site and the other in the BC site. Hex site as a stable site for the hydrogen molecules has not been previously reported. The $H_2$ molecule in the Hex site has the same stability as in the T site.

Interstitial molecules ($H_2$ in T and in Hex sites) and the $H_2^*$ complex are more stable than the neutral hydrogen atoms. Their energies with respect to the $H_{BC}^0$ configuration are -0.77 eV (for both $H_2$ in T and Hex sites) and -0.60 eV, respectively.

The neutral molecular complex $H_2^*$ is composed by two atoms: one in BC site the other in the nearest AB site. The charge density around these two atoms is homogenous without any preference for one of them.

In the T site $H_2$ interstitial molecules are more stable, after relaxation, along the [111] direction compared to the [110] direction as suggested elsewhere[34]. The H-H distance (0.77 Å) within the interstitial molecules in both T and Hex sites is almost equal to the distance of the free hydrogen molecule.

### 3.2.2 Migration and decomposition processes

The $H_2$ molecule migrates between two first order NN T sites. During migration, the molecule exhibits a rotation from [111] direction towards [100] direction in order to satisfy the steric constraints. During the migration, the distance between H atoms is increasing by 5 % in the midpoint site on its path between T sites. The migration barrier is 2.2 eV, five times higher than that of an isolate hydrogen atom (Figure 7).

Since the molecular interstitial is then, practically relative to atomic H, immobile we have considered its decomposition together with the decomposition of the other stable hydrogen dimers. Two decomposition processes have been considered:

(5) $$H_2(T) \rightarrow H_{BC}^+ + H_{AB}^-$$

(6) $$H_2^* \rightarrow H_{BC}^+ + H_{AB}^-$$

Decomposition of $H_2$ (Figure 8a), represented formally by equation (5), exhibits a quite simple barrier of 1.2 eV with an apparent shoulder at 0.6 eV. The molecule decomposes by the breaking of the H-H bond the migration of $H^+$ and $H^-$ at their respective sites within the octahedral Si cage.

Decomposition of $H_2^*$ (Figure 8b), represented formally by equation (6), exhibits a barrier of 1.5 eV. The decomposition of $H_2^*$ has been modelled as the simple migration of the $H^-(AB)$ from its initial position within the $H_2^*$ complex to a second order NN AB site.

# 4 Discussion

## 4.1 Comparison with other computational results

In terms of relative stabilities, when it comes to the neutral atoms series, our calculations show that the lowest configuration for the neutral atomic hydrogen is $H^0(BC)$ which is in perfect agreement with all previous works[21, 35-38]. The formation energy of $H^0(BC)$ with respect to the free hydrogen atom is -1.05 eV, a value in perfect agreement with Van de Walle and co-workers [20-22]. Our results are also in agreement with these works when it comes to the stable position of $H^+$, which has been found to be also the BC site.

While for $H^0$ and $H^+$ our results are in good agreement with most of the authors, however for the AB position, the minimum we found for the negative atoms series disagrees with recent works using plane wave DFT[20, 21]. These works claim that T is the stability site for $H^-$. This is due to the differences between our numerical schemes and that of the cited authors. Different numerical schemes can lead to different electronic densities and hence to different charge densities distributions. The short distance (only 0.5 Å) between the AB and T sites suggests that the difference is due to the spatial cut-off of the numerical orbitals employed by us.

The relative stabilities of $H_2$ and $H_2^*$ are in a very good respective agreement with the values of 0.87 eV and 0.60 eV for relative stabilities reported elsewhere[21]. One may see also that compared to the lowest neutral atomic species, the diatomic ones are more stable in a good agreement with the previous works[20, 34, 39-41].

## 4.2 Comparison with experimental results

Despite the minor disagreement between our DFT results and other authors, related to $H^-$, our calculations are predicting two stability sites that are found also by µSR: BC site for $H^+$ and a site between the AB and T sites for $H^-$. Moreover, the charge of the hydrogen atoms in these sites is the respective agreement with the results predicted by µSR[42-46] and DLTS[47, 48].

A remarkable result reported within this work is the confirmation of the negative – U behaviour of H atoms in the Si lattice by computational means. Through our calculations we have successfully calculated the U constant corresponding to the process $2H^0 \xrightarrow{U} H^+ + H^-$. The value U = 0.32 eV is in very good agreement with the experimental value[30, 31] of U = 0.36 eV. However the position of the Fermi level in the gap within the gap is different than that predicted by the experiment. This is caused by the well-known issue of the DFT-based computational frameworks of failing to predict well the band gap and the levels in the gap.

Finally, another successful result is the migration barrier of hydrogen atoms which is linking in fact two distinct experimental facts:

1. the hydrogen atoms are the migrating species in Si at high, room and cryogenic temperatures[49].
2. the hydrogen atoms migrating in Si lattice are charged, their charge sign depending on the Fermi level position within the band gap: $H^+$ is the migrating species in p-type Si and $H^-$ is the migrating species in the *n*-type Si.

The value of the activation energy for hydrogen diffusion in the barely damaged Si crystal is the same (0.46 eV) for both high and cryogenic temperatures[49, 50].

## 4.3 Hydrogen diffusion model

Our numerical scheme is validated by the qualitative and quantitative agreement between our results with previous works both computational and experimental as it has been explained in detail in the previous sub-sections

The fact that at 0 K (the temperature to which DFT calculations are performed) the migration barrier of atoms is significantly smaller to that of molecules shows that these species are regulating and limiting the hydrogen transport within the silicon lattice.

The migrating species are hydrogen ions that can bear either positive or negative charge. Though the migration barrier is identical for both migrating species, the mechanisms for their migration are different. Positive ions migrate between two equivalent NN BC sites while the negative ions migrate between second order neighbours AB sites. Hence one may conclude that negative ions migrate faster than the positive ions, the length of their elementary migration path being larger than that of positive ions.

It is known however that in n-type Si[37, 51, 52] the dominating migration species is $H^-$, while in the p-type Si[53, 54] the dominating migration species is $H^+$ due to their respective stabilities given by the position of the Fermi level in these two doping cases.

At low temperatures, the hydrogen atoms may form diatomic structures during their migration process, given that these later structures are more stable than the former ones. The recombination barriers between two hydrogen ions (one positive and one negative) are relatively low, of the same order of magnitude as the migration barrier itself: 0.2 eV to form an interstitial $H_2$ and 0.4 eV to form a $H_2^*$. Thus, one can see that the formation of the interstitial molecules is favoured with respect to the formation of $H_2^*$. The molecule formation is limited by the capture probability at the T site. Hence the limitation is linked to the probability of two ions baring different charges to meet at the T site.

Increasing the temperature would lead to the release of new hydrogen ions through the decomposition of $H_2$ and $H_2^*$ structures. The decomposition barrier of $H_2$ is slightly smaller to that of the $H_2^*$: 1.2 eV compared to 1.5 eV.

The diffusion of H in Si is energetically defined by the diffusion activation energy. If we consider the case that all hydrogen is atomic only, then the diffusion activation energy is the migration energy of the atomic species. If we consider the case that all H is under dimmer forms, then first the bond has to be broken before diffusion can take place. Therefore one can define the diffusion activation energy which equals the energy required for dissociation plus the atomic migration barrier.

Hence, the presence of molecules in the H implanted Si leads to higher diffusion activation energies. Higher diffusion energies have been measured experimentally and they have been attributed to hydrogen trapping either at molecule sites or to the dangling bonds[55-59]. In low vacancy concentration silicon, our results favour the scenario of molecule formation and then its decomposition.

Both of dimer decomposition barriers are smaller to the migration barrier of $H_2$ which means that these decomposition processes are more probable than the migration of $H_2$. The high barrier of migration of the interstitial molecule show that these ones do not participate as migrating species to the diffusion process. However, experimental analysis of IR data allows inferring that $H_2$ diffuses at temperatures higher than 300°C[2, 3], while our calculations show that actually the $H_2$ molecule decomposes before any migration process.

## 5   Conclusions

In this paper we have presented DFT calculation of the structure, the energetics and of the kinetic properties of hydrogen defects: hydrogen ions, hydrogen molecules and hydrogen diatomic defect $H_2^*$. Our results show that in silicon, the dominant migration species are hydrogen ions either positive or negative. The migration barriers of these defects are identical while their elementary migration paths are different. These ions may form diatomic structures that are energetically more stable. With the formation of these diatomic structures the hydrogen transport halts. Increasing of the temperature leads to the decomposition of these diatomic structures that are practically immobile with respect to the H ions.

# Figures

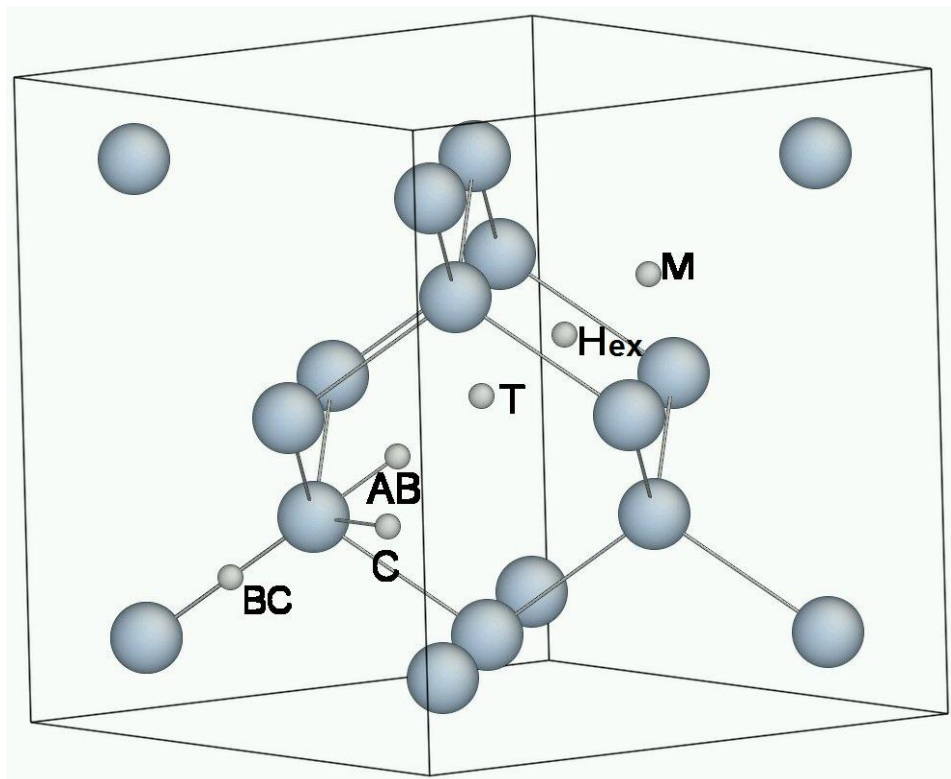

Figure 1 The positions of various interstitial sites with respect to the [111] axis.

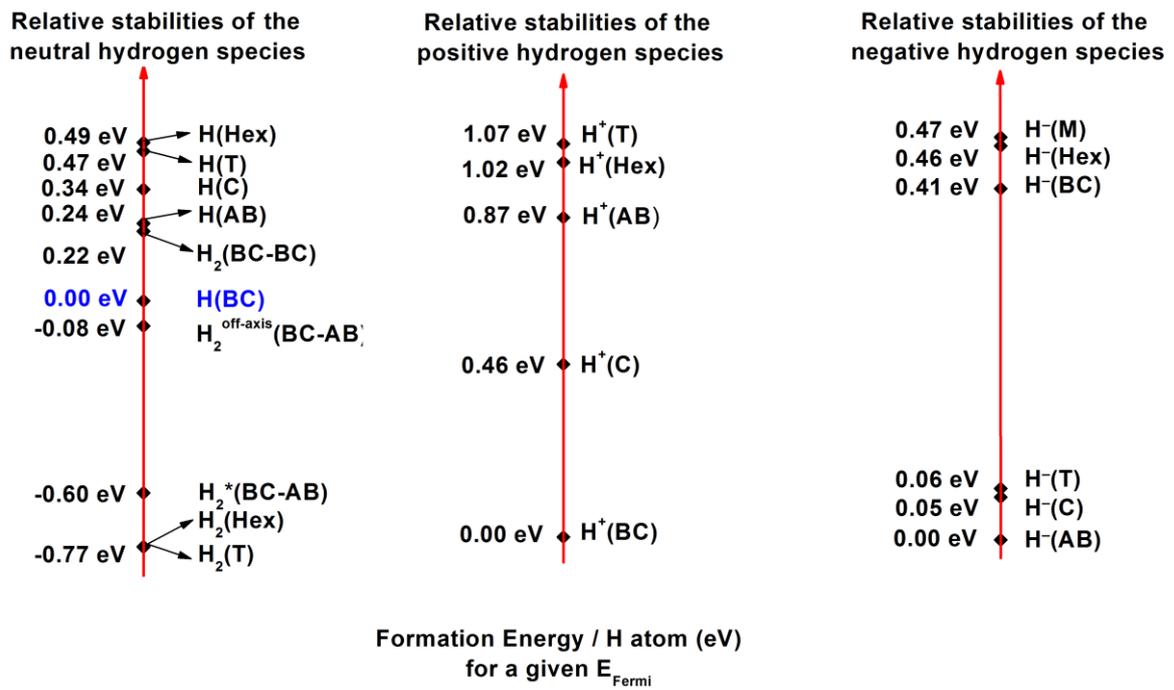

Figure 2 Stability of various atomic and molecular configurations (a) neutral species stability with respect to the energy of the neutral hydrogen atom in BC site. Two pairs of structures are overlapping from energetic view point: the H(M) coincides with H(BC) and $H_2$(H) is only 0.01 eV higher than $H_2$(T); (b) the stability of positive monatomic species with respect to the minimum energy species for this series, $H^+$(BC); (c) the stability of negative monatomic species with respect to the minimum energy species for this series, $H^-$(AB).

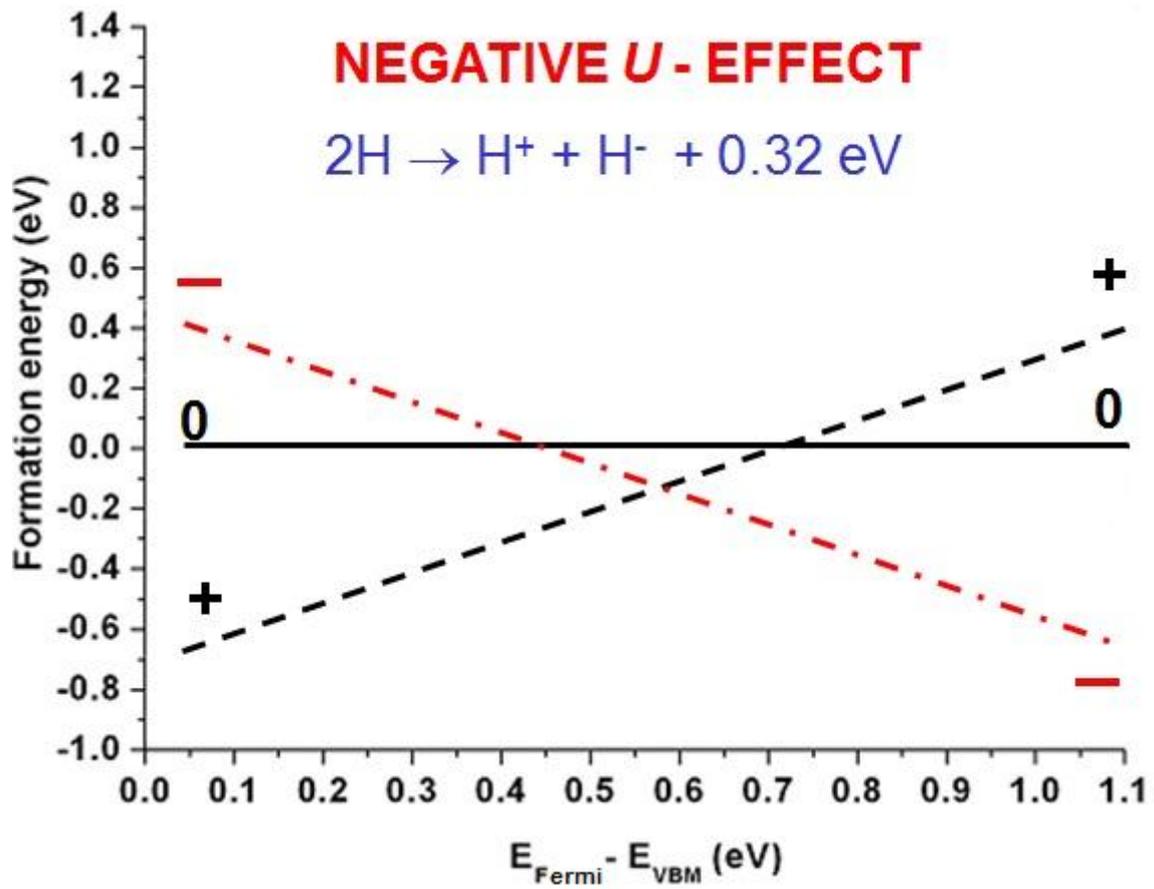

Figure 3 Formation energy of charged atomic interstitials in their respective minimum energy sites as a function of the Fermi energy. The horizontal line represents the energy of $H^0$ in the BC site, the left-tilted and the right-tilted lines are representing the energy of the positive and negative hydrogen ions in the BC and AB sites respectively.

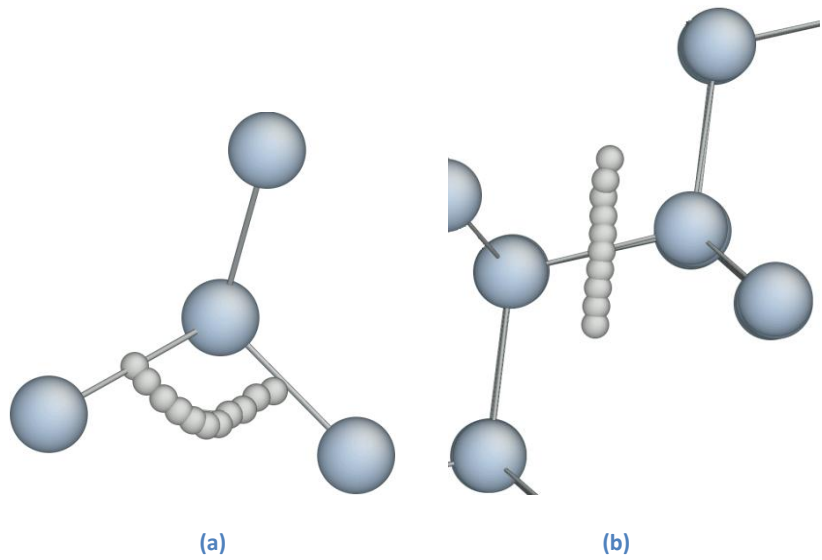

(a)                 (b)

Figure 4 Migration of hydrogen atoms within Si lattice : (a) positive ions between 1$^{st}$ order nearest-neighbour (NN) BC sites; (b) negative ions between 2$^{nd}$ order NN AB sites.

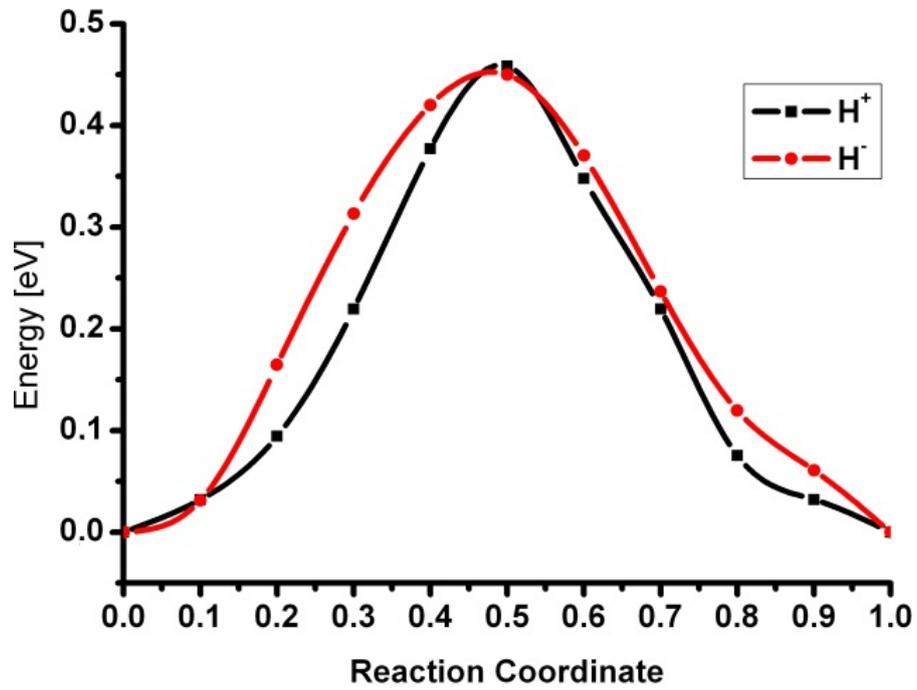

Figure 5 Migration of hydrogen atoms within the Si lattice: in black, the migration barrier of H$^+$ between two first NN BC sites and in red the migration barrier of H$^-$ between two second order NN AB sites.

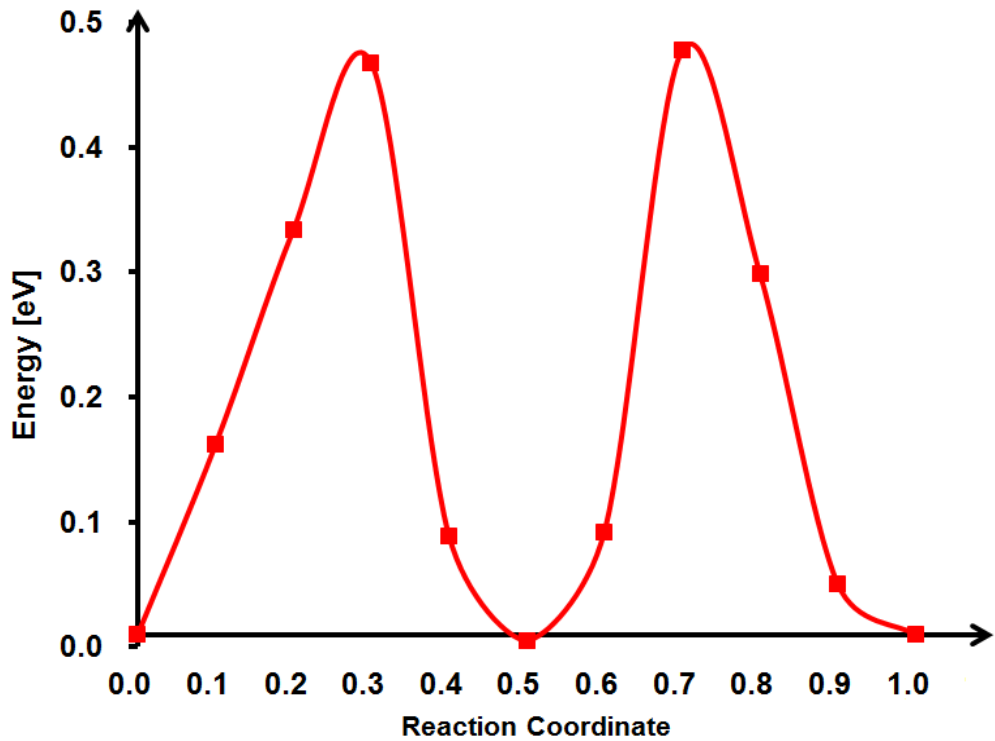

Figure 6 The migration barrier of H$^+$ between two second NN BC sites composed by two elementary barriers between first NN BC sites.

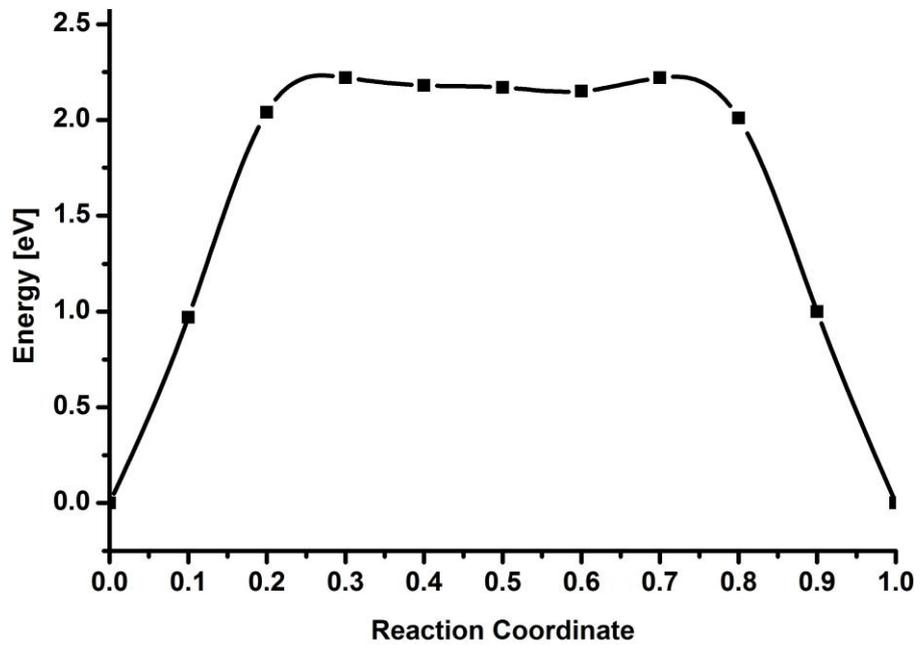

Figure 7 The migration barrie of the hydrogen molecule between two next-nearest neighbour T sites within the Si lattice.

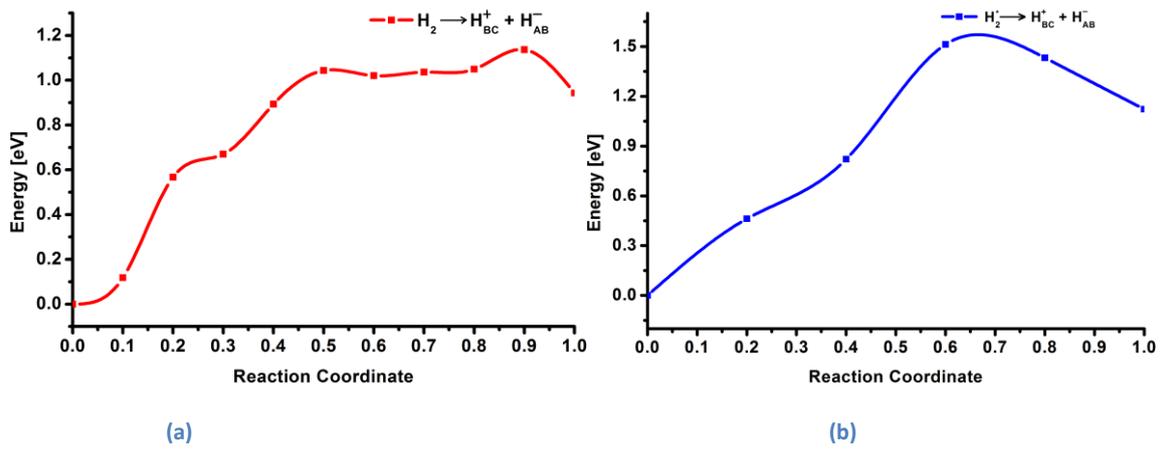

(a)    (b)

**Figure 8** Decomposition barriers of the two stable hydrogen dimers in silicon: (a) $H_2$ decomposition represented by equation (5) and (b) $H_2^*$ decomposition represented by equation (6).